**Title**: Harnessing Big Data and Artificial Intelligence to Study Plant Stress

**Authors**: Eugene Koh[1], Rohan Shawn Sunil[1], Hilbert Yuen In Lam[1], Marek Mutwil[1*]

1 School of Biological Sciences, Nanyang Technological University, Singapore, Singapore

## Abstract

Life finds a way. For sessile organisms like plants, the need to adapt to changes in the environment is even more poignant. For humanity, the need to develop crops that can grow in diverse environments and feed our growing population is an existential one. The advent of the genomics era enabled the generation of high-throughput data and computational methods that serve as powerful hypothesis-generating tools to understand the genomic and gene functional basis of stress resilience. Today, the proliferation of artificial intelligence (AI) allows scientists to rapidly screen through high-throughput datasets to uncover elusive patterns and correlations, enabling us to create more performant models for prediction and hypothesis generation in plant biology. This review aims to provide an overview of the availability of large-scale data in plant stress research and discuss the application of AI tools on these large-scale datasets in a bid to develop more stress-resilient plants.

*Corresponding author: Marek Mutwil School of Biological Sciences Nanyang Technological University 60 Nanyang Drive 637551 Singapore

Email: mutwil@ntu.edu.sg

## Introduction

The intricate relationship between plants and their environment forms the bedrock of ecological balance and sustenance. In the face of evolving climatic patterns, burgeoning populations and environmental degradation, understanding how plants respond to stress has become a matter of paramount significance (Verslues et al., 2023). Environmental stress, encompassing both abiotic and biotic factors, poses formidable challenges to plant life, affecting growth, productivity and overall ecological resilience. The repercussions of these stressors extend beyond individual plants, influencing entire ecosystems and, consequently, the global environment (Intergovernmental Panel on Climate Change (IPCC), 2023). As we navigate an era marked by climate change and global ecological shifts, the ability to discern how various plant species adapt and thrive—or falter—under diverse stress conditions is pivotal. This understanding not only informs sustainable agricultural practices but also holds the key to preserving biodiversity, ensuring ecosystem resilience, and ultimately, securing our planet's environmental well-being.

While stress is intensively studied by molecular biologists who have already uncovered multiple genes and biological processes underpinning stress resilience mechanisms (Saijo & Loo, 2020; Zhang et al., 2022; J. K. Zhu, 2016), computational biologists can also use advances in large-scale data generation to form powerful predictive models about gene function (Rhee & Mutwil, 2014). In this review, we will cover how large-scale data can be used to advance our understanding of stress resilience and fuel powerful AI approaches able to predict stress resilience mechanisms.

*Kingdom-level overview of genomic and transcriptomic plant stress studies*

Genomic and transcriptomic data are invaluable to understanding gene functions and how organisms respond to the changing environment. Databases such as the Sequence Read Archive (SRA) and Gene Expression Omnibus (GEO) play pivotal roles in conducting meta-analyses of environmental stress studies across the plant kingdom (Edgar, 2002; Leinonen et al., 2011). These repositories serve as invaluable reservoirs of high-throughput sequencing data, transcriptomic profiles, and other omics datasets, providing researchers access to a wealth of information derived from diverse experimental setups (Julca et al., 2023). The GEO database was launched by NCBI in 2000 as a repository for high-throughput gene expression data (Edgar, 2002), which were dominated initially by data from DNA microarray experiments. Today, abundant omics data from protein or tissue arrays, genome methylation, genome binding/occupancy, protein profiling, chromosome conformation, and genome variation/copy number studies can also be found in GEO (Clough & Barrett, 2016).

The SRA was established as a public repository for next-generation sequence data as a part of the International Nucleotide Sequence Database Collaboration (INSDC) (Karsch-Mizrachi et al., 2012), that acts as a nexus collating raw next-generation genomic and transcriptomic sequencing data, and is closely integrated with other large databases such as the ArrayExpress at EBI (http://www.ebi.ac.uk/arrayexpress) (Parkinson et al., 2011) and DDBJ Omics Archive (http://trace.ddbj.nig.ac.jp/dor) (Kodama et al., 2012) hosted by the European Bioinformatics Institute (EBI) and DNA Data Bank of Japan (DDBJ). Both databases share significant overlaps, with new accessions in GEO automatically and simultaneously uploaded to SRA as well, but not vice versa. Besides the biological data present in these databases, plentiful metadata also exists in these databases which provide contextual information on the purpose of the studies, or detailed information on specific treatment conditions, plant organ type, species / genotype / cultivar variations, plant age / developmental stage, sample geographical origin and others. These metadata can be further analysed to identify specific research trends of scientific interest, or which may be useful for funding bodies and policymakers.

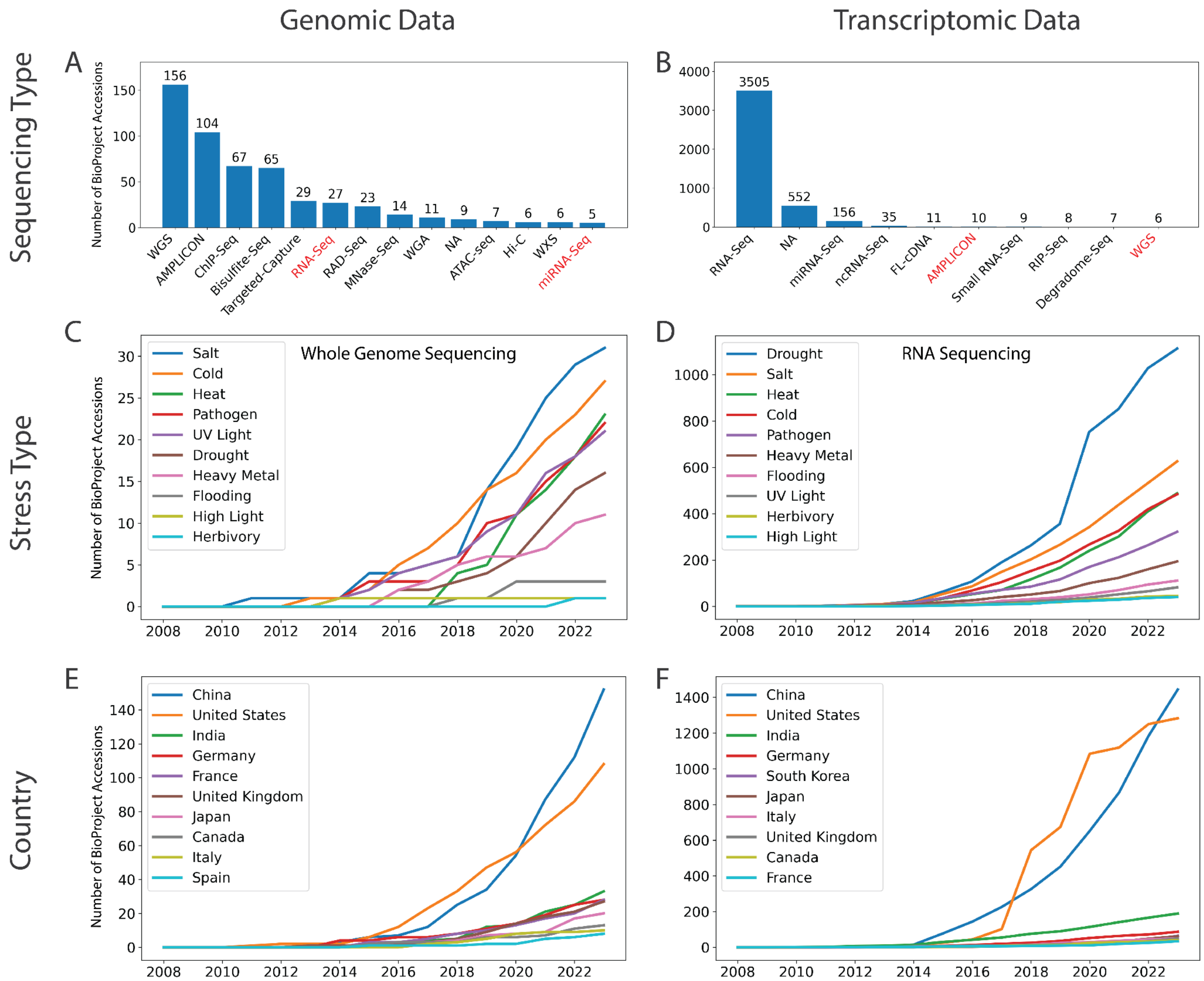

**Figure 1. Overview of stress-related BioProject IDs present in the SRA database for genomic and transcriptomic studies.** (A, B) After filtering for stress-related BioProject IDs based on our keywords, BioProjects were classified based on the sequencing type used, only displaying those with >5 BioProjects. Misannotated sequencing types are labelled in red. (C, D) The cumulative number of BioProjects in SRA over the years were also classified by stress type, and (E, F) country of origin. Figure legends show the respective ranks of the cumulative total number of BioProjects.

In order to derive a global overview of the data available from stress studies across the plant kingdom, we began by querying for the term 'Viridiplantae' within the SRA database, which encompasses two major clades: Streptophyta, which contains streptophyte algae and land plants (embryophytes), and Chlorophyta, which includes all remaining green algae. We observed that the SRA houses a total of 558,292 transcriptomic and 1,387,717 genomic accessions for Viridiplantae at the time of writing (May 2024), and we performed a preliminary grouping of this data into their respective unique BioProject accessions. The BioProject database is a comprehensive and centralised repository managed by the National Center for Biotechnology Information (NCBI) that serves as a platform for the organisation and dissemination of information related to biological research projects (Barrett et al., 2012). Each BioProject entry includes detailed information about the project's objectives, experimental design, and data sources. The metadata available at this step allowed us to filter for experiments conducted in stress studies encompassing 10 specific stress types we had defined: high light, heat, cold, UV light, salt, drought, heavy metal, pathogen, flooding and herbivory. This was achieved by parsing the metadata with keywords specific to each of the 10 stress types. In total, we retrieved 4323 and 592 stress-related BioProjects for transcriptomic and genomic sets, respectively. By inspecting the metadata, we could determine the types of experimental procedures performed to obtain the respective genomic and transcriptomic data. As expected, whole genome sequencing (WGS) and RNA-Seq were the most common methods used to obtain genomic and transcriptomic data (Figure 1). For genomic data, we observed a significant number of BioProjects which were tagged with epigenomic type studies performed with ChIP-Seq, BiSulfite-Seq, ATAC-Seq etc, which would highlight the studies characterising the various stress-induced epigenomic changes. The WGS accessions obtained are possibly part of larger studies for previously unsequenced species. For transcriptomic data, we observed that microRNA (miRNA) sequencing, noncoding RNA (ncRNA) sequencing, full-length cDNA (FL-cDNA) sequencing and small RNA sequencing were the next most popular transcriptomic methods. We noticed that there was a significant number of unannotated or incorrectly annotated entries, for instance, RNA-Seq or miRNA-Seq (coloured in red) being tagged in genomic data or unclassified samples (NA), WGS and Amplicon sequencing being tagged in transcriptomic data.

*Drought, salt, heat, cold and pathogen stress dominate the stress study landscape*

Next, to understand which stresses are most studied, we plotted the number of unique BioProject IDs annotated with specific stress-related keywords. We observed that drought, salt, heat, cold and pathogen stress were the top five most common stress types for genomic (Figure 1C) and transcriptomic (Figure 1D) data. Of note is the observation that the accumulation of genomic and transcriptomic data present in the SRA began around 2010-2012, when next-generation sequencing technologies such as RNA-Seq came to the fore, supplanting the previous generation of transcriptomic technologies such as microarrays. Transcriptomic data derived from microarrays were previously stored in repositories such as ArrayExpress and the Gene Expression Omnibus (GEO) databases but are not represented in our chart.

*Two nations produce >80% of all available data*

The metadata present in our dataset also allows us to plot the progress of scientific investment into plant stress research in a country-specific manner, using the submitting institutions as a proxy (Figure 1). From the charts, we observed that the United States, China and India were the top three drivers of plant stress research, with the United States having led for close to a decade but has been recently overtaken by China. The interest shown by both superpowers in understanding the mechanisms involved in plant stress signalling is an encouraging sign for the field and is extremely timely considering the looming effects of climate change affecting global food security. Other members include developed countries in Europe and Asia, such as Germany, France, the United Kingdom, Italy, Spain, Japan and South Korea. Of note are the obvious absences of developing countries and those from the global South, which hints at a potential widening divide between the rich and poor nations of the world in preparation for the incoming climate shifts.

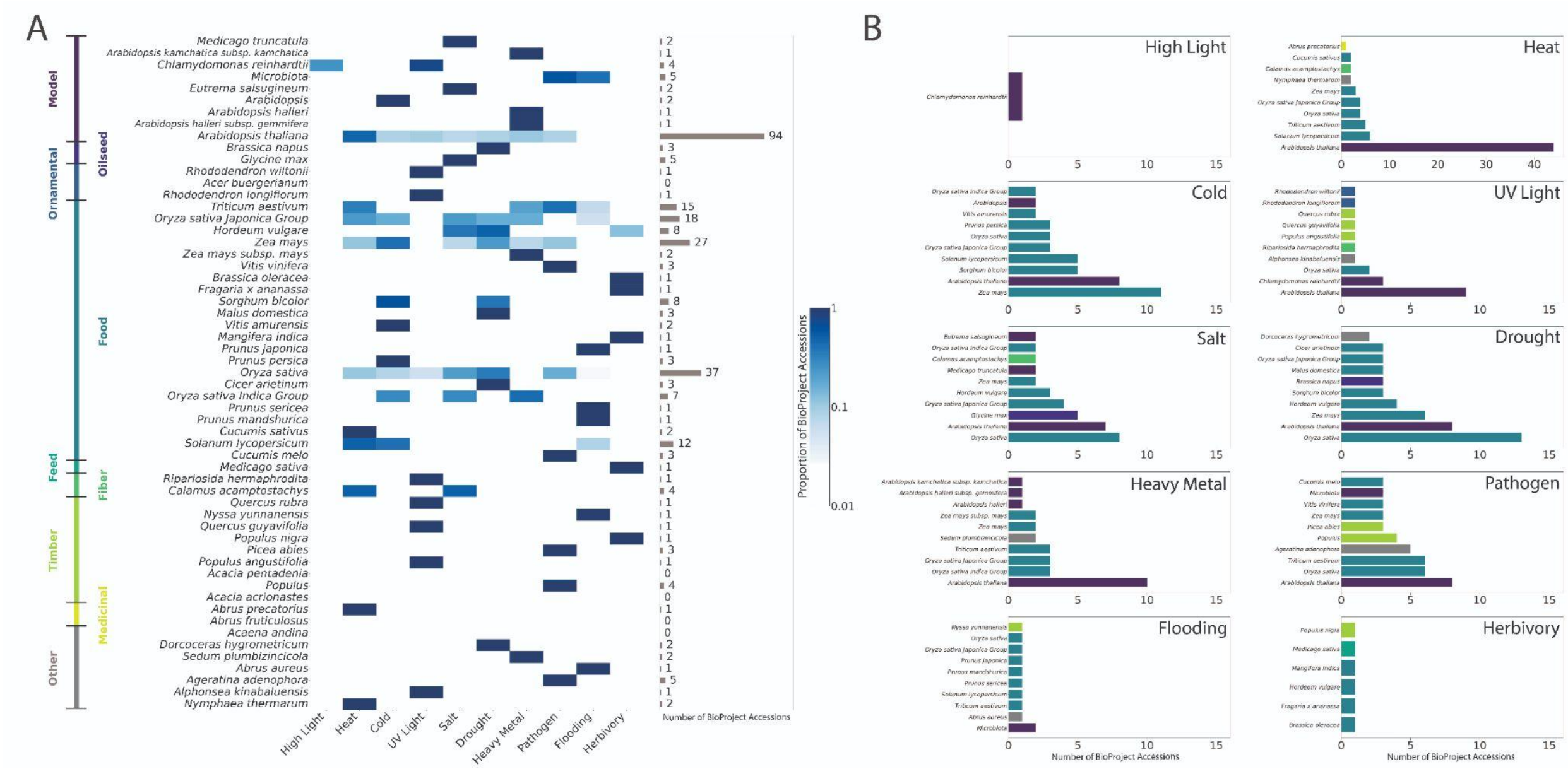

**Figure 2. Distribution of top 10 species per stress studied across stress-related BioProject IDs for genomic data**. The counts of different species studied across the BioProject IDs for each stress type are represented in a (A) log-scale heat map normalised within each species, and in (B) bar charts showing the top few species. Some stresses contained <10 species studied.

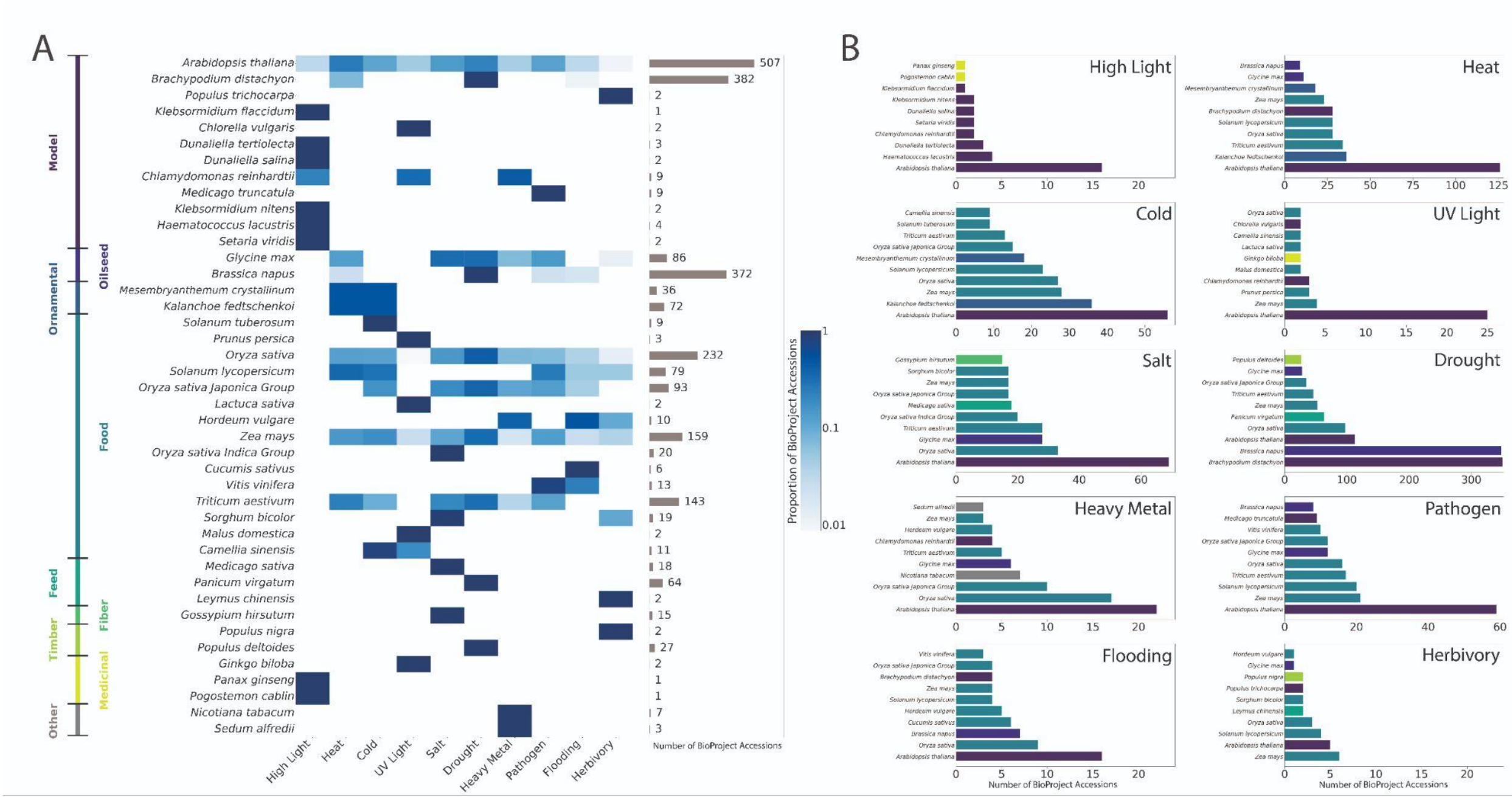

**Figure 3. Distribution of top 10 species per stress studied across stress-related BioProject IDs for transcriptomic data**. The counts of different species studied across the BioProject IDs for each stress type are represented in a (A) log-scale heat map normalised within each species, and in (B) bar charts showing the top few species.

*Species- and stress-specific research trends*

Next, we investigated how the various species were used for stress research. To that end, we identified the top 10 species per stress type that were studied by genomic (Figure 2) and transcriptomic (Figure 3) approaches and divided the species based on their uses (e.g., crop, model, ornamental). Overall, food crops were most comprehensively represented (27 out of 82 species), followed by model plants (18 out of 82) and timber crops (10 out of 82). For transcriptomic data, heat, cold, salt, drought and pathogen stress were the most abundant (>100 BioProject accessions), followed by UV, heavy metal and flooding stress (50-100 BioProject accessions), and high light and herbivory stress (<50 BioProject accessions) (Figure 3). It was unsurprising to note that the model plant *Arabidopsis thaliana* dominated for most stress types, with the exception of drought stress, where *Brachypodium distachyon* and *Brassica napus* led the field; and herbivory, where *Zea mays* came in slightly ahead of Arabidopsis (Figure 3). Since Arabidopsis contained the highest number of samples, we decided to identify species coming in second to fifth places to reveal current research efforts. We observed that the crop plants *Triticum aestivum*, *Oryza sativa*, *Solanum lycopersicum*, *Glycine max* and *Nicotiana tabacum* were strongly represented in the majority of the stress types for genomic (Figure 2) and transcriptomic (Figure 3) data. Of interest was the presence of *Kalanchoe fedtschenkoi* in second place in heat/cold stress and *Panicum virgatum* in fifth place in drought stress, but with still more than 50 BioProject accessions. Kalanchoe is a common ornamental houseplant (Smith et al., 2019), while Panicum is used as cattle feedstock (Griffin & Jung, 2019) or also as ground cover to prevent soil erosion (E. Wang et al., 2020). The position of Kalanchoe in second place for heat/cold stress was unexpected, which prompted us to examine the source data more closely. We found that the observed 'overrepresentation' of Kalanchoe was due to individual accessions from the same submitting institution being assigned with unique BioProject IDs in SRA. This is a distinct example of the difficulty of standardising and curation of metadata due to inadvertent human error and other inconsistencies. Nevertheless, the rise in atypical species such as these is an encouraging sign that scientific knowledge, interests and resources are percolating through the system and leading to more translational discoveries.

High light research was conducted to a large extent on Arabidopsis (Figure 2-3), but this stress was surprisingly not studied in crop plants, and only a few datasets were present in other model organisms. On the other hand, UV research was also dominated by Arabidopsis, but with several common crop plants like *Prunus persica*, *Malus domestica*, *Camellia sinensis* and even *Ginkgo biloba*, known for its medicinal properties, also in the mix. Climate change can come with changing rainfall patterns and consequently decreased cloud cover, while geographical locations at risk of desertification are also exposed to such stresses (Barnes et al., 2023). High intensities or prolonged duration of visible or UV light can cause significant Reactive Oxygen Species (ROS) production, resulting in damage to various biomolecules such as DNA/RNA, proteins and lipids, resulting in cellular dysfunction and the accumulation of mutations (Quiroz et al., 2023). We believe that increasing focus on these relatively neglected areas of research would be pivotal in understanding the chronic biological effects of these stresses.

For genomic data, other than heat stress, which had >50 BioProject accessions, we found <30 BioProject accessions for the remaining stress types, with there being even fewer (<5 BioProject accessions) for high light, flooding and herbivory stress in particular (Figure 2). Overall, this showcases considerably fewer stress-related genomic studies as compared with those utilising transcriptomic data. From a fundamental standpoint, this does make sense, as changes in gene expression would be one of the main focuses for studies investigating the effect of various stresses on plants. Again, *Arabidopsis thaliana* was a key contributor in seven of the stress types in terms of genomic data, with *Zea mays* being the most studied for cold stress and *Oryza sativa* for salt and drought stress.

*Online resources providing analyses of large-scale data*

A vast variety of bioinformatic tools were developed that utilise the genomic, epigenomic, transcriptomic, proteomic, and other '-omic' data (Cantó-Pastor et al., 2021). For genomics, the 1001 Genomes Project capturing natural variation data allows the exploration of 462 phenotypes across 1496 accessions in *Arabidopsis thaliana* (Alonso-Blanco et al., 2016). One of the most widely used portals, TAIR (http://www.arabidopsis.org, (Reiser et al., 2017)), provides detailed information about gene functions and maintains a 'super-portal' to keep track of and categorise various Arabidopsis tools (https://conf.arabidopsis.org/display/COM/Resources). Databases such as Ensembl Plants (Kersey et al., 2018), PLAZA (Van Bel et al., 2018) and PANTHER (Mi et al., 2010), allow the exploration of gene families and gene trees. Epigenomic DNA modifications from numerous sequencing experiments can be viewed in the EPIC-CoGe Browser (Nelson et al., 2018) and the 1001 Epigenomes Browser (Kawakatsu et al., 2016), giving the researchers unprecedented means to study the epigenetic regulation of genes.

Transcriptomic tools allow the analysis of expression profiles across publicly available transcriptomic data comprised of RNA-sequencing and microarrays (Julca et al., 2023), and are accessible through sites such as the eFP browser (http://bar.utoronto.ca/efp/cgi-bin/efpWeb.cgi)

(Winter et al., 2007)), Arabidopsis RNA-Seq database (http://ipf.sustech.edu.cn/pub/athrna, (Zhang et al., 2020)), Plant Single Cell RNA-Sequencing Database (https://www.zmbp-resources.uni-tuebingen.de/timmermans/plant-single-cell-browser/; (Ma et al., 2020)),CoNekT (https://conekt.sbs.ntu.edu.sg/, (Proost & Mutwil, 2018)) and Plant Expression Omnibus (https://expression.plant.tools/, (Koh et al., 2024)). Since genes with similar expression patterns (co-expression) can be functionally related, several tools that explore co-expression networks for Arabidopsis and other species are available. The tools include ATTED-II (https://atted.jp/, (Obayashi et al., 2018)), CoNekT (https://conekt.sbs.ntu.edu.sg/, (Proost & Mutwil, 2018)), Expression Angler (https://bar.utoronto.ca/ExpressionAngler/, (Austin et al., 2016)), and others.

Proteomic tools range from methods that study protein subcellular localisation (https://version4legacy.suba.live/, (Hooper et al., 2017)), protein modifications, such as phosphorylation, acetylation, methylation, nitrosylation, ubiquitination and glycosylation (http://p3db.org, http://www.psb.ugent.be/PlantPTMViewer, (Willems et al., 2019; Yao et al., 2014)). Since interacting proteins are also likely functionally related, tools to visualise protein-protein interactions, such as BAR's Arabidopsis Interactions Viewer 2 (https://bar.utoronto.ca/interactions2/, (Dong et al., 2019)) and AtPID (http://www.megabionet.org/atpid/, (P. Li et al., 2011)). Finally, with the advances in AI, accurate protein structures can be predicted (https://alphafold.ebi.ac.uk, (Jumper et al., 2021)), opening new possibilities for structure-function studies.

There are a number of plant stress specific databases which have been published, which are based on extensive manual curation, and cater to a variety of species and stress types. A majority are transcriptomics-based (Plant stress RNA-seq Nexus (PSRN) (J.-R. Li et al., 2018), Plant Environmental stress transcript database (Balaji et al., 2006), Plant Resistance Genes database (PRGdb 4.0) (Calle García et al., 2022), DroughtDB (Alter et al., 2015), Stress Responsive Transcription Factor Database (STIFDB V2.0) (Naika et al., 2013), CerealESTDb (Kumar et al., 2022)) or genomic databases (HEATSTER (Berz et al., 2019), Rice Stress-Resistant SNP Database (Tareke Woldegiorgis et al., 2019)), but more specialised databases focusing on specific non-coding RNAs such as (PncStress database (Wu et al., 2020), CropCircDB (K. Wang et al., 2019)) or alternative splicing events (PlaASDB (Guo et al., 2023)) also exist in the literature. Of special mention is the Stress Knowledge Map, which provides information of specific protein-protein interactions, protein-DNA interactions, small RNA-transcript interactions and enzymatic transformations of metabolites in the form of knowledge graphs (Bleker et al., 2024). These databases provide a useful resource for the community and are summarised in Table 1.

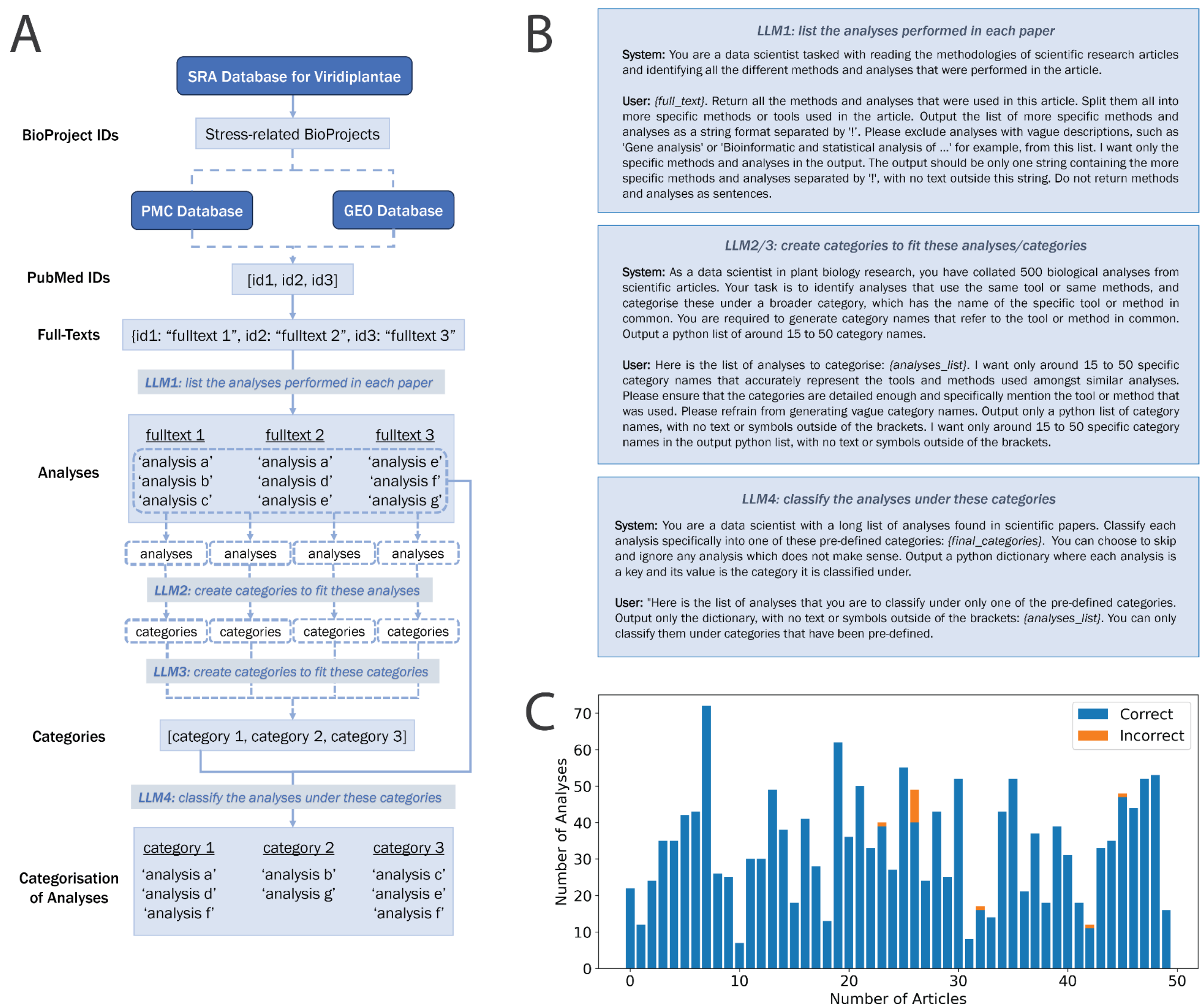

**Figure 4. Categorisation of analyses performed across transcriptomic studies.** (A) Stress-related BioProject accessions with transcriptomic data were used to obtain their associated research articles. All the performed analyses that were stated in the articles were classified into broader supergroups using a series of prompts with the OpenAI API, as further detailed in (B). The accuracy of LLM1 was validated with 50 random articles by manually cross-checking whether each analysis generated was mentioned in the article. The percentage of correct analyses generated by LLM1 can be seen in (C), giving a mean accuracy value of 99.3%.

*Large language model-assisted literature mining and data analysis*

While genomic and transcriptomic approaches are invaluable to studying plant stress resilience mechanisms, these approaches are typically used in combination with other methods. To better understand how the different methods are combined in research papers, we performed a large language model (LLM)-based analysis of methods appearing in the papers that generated the stress-related genomic and transcriptomic data. The advent of AI tools provides a unique opportunity to rapidly mine the burgeoning scientific literature, which was previously performed through arduous manual curation. Recently, we deployed the text mining capacities of Generative Pre-trained Transformer (GPT) to process over 100,000 plant biology abstracts, revealing nearly 400,000 functional relationships between a wide array of biological entities - genes, metabolites, tissues, and others, with remarkable accuracy of over 85% (Fo et al., 2023), and have further applied this tool across multiple species, such as yeast (Arulprakasam et al., 2024). Here, we used GPT to perform a relational analysis of the experimental tools and methods commonly used in the study of plant stress in a bid to create a knowledge network for the training of plant biology specific models.

A typical research paper comprises several methods that are used to generate data to answer a scientific question. To explore which methods are typically employed in articles that use high throughput data and which methods tend to be used together, we performed a LLM-based analysis of ~2,000 articles associated with the total set of BioProject accessions discussed in the previous section for transcriptomic data, representing all plant stress studies recorded in the SRA (Figure 4A). To this end, we queried the PubMed Central (PMC) and GEO databases where these BioProject accessions were referenced and obtained PubMed IDs for downloading the relevant research articles. Not all BioProject accessions have an associated PubMed ID, so in total, 2041 articles (1769 full-text, 272 abstract only) were downloaded from a starting list of 4323 unique stress BioProject IDs. Using the OpenAI API, we performed four LLM queries to help us parse and categorise the various research methods used in the plant stress literature (Figure 4B). In our workflow, LLM1 was tasked to recognise and list the various experimental and bioinformatic methods performed in the 'Methods' section of our list of full-text articles. As LLMs are prone to hallucinating, we manually checked the accuracy of the LLM output for 50 random papers against their respective texts to ensure that no superfluous terms were listed. This quality control check showed that LLM1 had an accuracy value of 99.3% at its task (Figure 4C). LLM2 and LLM3 were designed to take the output lists and iteratively categorise them into groups and supergroups, respectively. Finally, LLM4 was tasked to group the original full-text derived lists into the supergroups obtained by LLM3. Within the OpenAI API, the "temperature" parameter dictates the variability of the output. Of the 4 LLMs, we adjusted the temperature of LLM1 and LLM4 to 0 to reduce hallucination but provided LLM2 and LLM3 (responsible for supergroup assignment) with a temperature of 0.5 to allow greater flexibility and creativity, as higher temperature values introduce more variation. In total, LLM1 identified 50,560 experimental approaches (e.g., alignment of protein sequences using MAFFT, Screening of differentially expressed genes using DESeq2), LLM2 combined every 500 approaches into 15-50 groups (e.g., microRNA (miRNA) Analysis, Read Quality Control - FASTX), and LLM3 further condensed these groups into 106 supergroups. Finally, LLM4 reclassified the 50,560 approaches into the 106 supergroups.

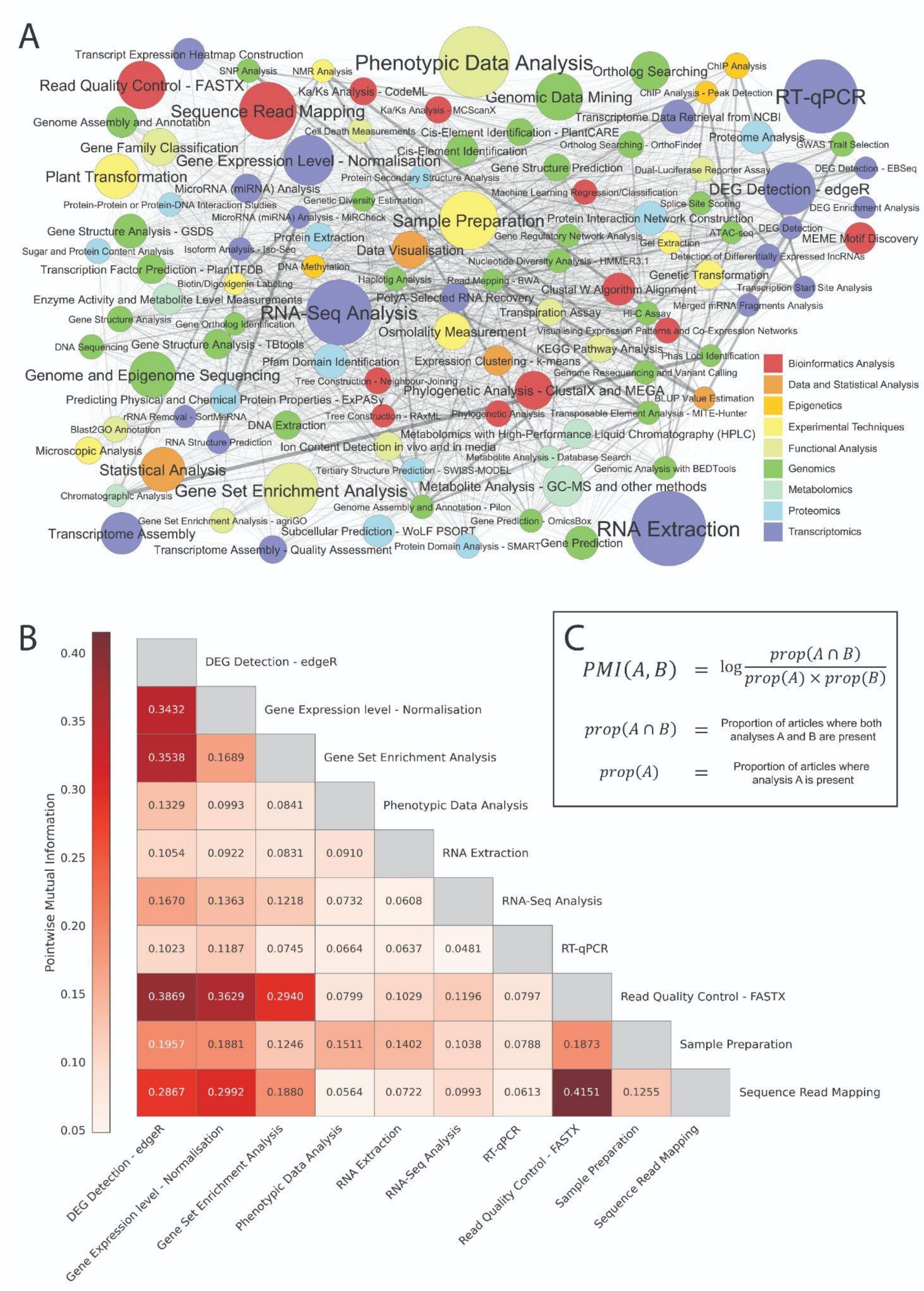
A
Transcript Expression Heatmap Construction
SNP Analysis
NMR Analysis
Phenotypic Data Analysis
Ortholog Searching
ChIP Analysis
Read Quality Control - FASTX
Ka/Ks Analysis - CodeML
Genomic Data Mining
ChIP Analysis - Peak Detection
RT-qPCR
Genome Assembly and Annotation
Sequence Read Mapping
Ka/Ks Analysis - MCScanX
Transcriptome Data Retrieval from NCBI
Cell Death Measurements
Cis-Element Identification - PlantCARE
Proteome Analysis
Gene Family Classification
Ortholog Searching - OrthoFinder
GWAS Trait Selection
Cis-Element Identification
Gene Expression Level - Normalisation
Plant Transformation
Gene Structure Prediction
Dual-Luciferase Reporter Assay
DEG Detection - EBSeq
Protein Secondary Structure Analysis
MicroRNA (miRNA) Analysis
Machine Learning Regression/Classification
DEG Detection - edgeR
Genetic Diversity Estimation
Protein-Protein or Protein-DNA Interaction Studies
Splice Site Scoring
DEG Enrichment Analysis
MicroRNA (miRNA) Analysis - MiRCheck
Sample Preparation
Protein Interaction Network Construction
Gene Structure Analysis - GSDS
ATAC-seq
DEG Detection
Protein Extraction
Gene Regulatory Network Analysis
MEME Motif Discovery
Gel Extraction
Sugar and Protein Content Analysis
Isoform Analysis - Iso-Seq
Data Visualisation
Detection of Differentially Expressed lncRNAs
Nucleotide Diversity Analysis - HMMER3.1
Transcription Factor Prediction - PlantTFDB
DNA Methylation
Genetic Transformation
Haplotig Analysis
Read Mapping - BWA
Biotin/Digoxigenin Labeling
Clustal W Algorithm Alignment
Transcription Start Site Analysis
Enzyme Activity and Metabolite Level Measurements
PolyA-Selected RNA Recovery
Merged mRNA Fragments Analysis
Gene Structure Analysis
RNA-Seq Analysis
Transpiration Assay
Hi-C Assay
Gene Ortholog Identification
Osmolality Measurement
Visualising Expression Patterns and Co-Expression Networks
DNA Sequencing
Gene Structure Analysis - TBtools
KEGG Pathway Analysis
Phas Loci Identification
Pfam Domain Identification
Expression Clustering - k-means
Genome and Epigenome Sequencing
Genome Resequencing and Variant Calling
Tree Construction - Neighbour-Joining
Phylogenetic Analysis - ClustalX and MEGA
Predicting Physical and Chemical Protein Properties - ExPASy
BLUP Value Estimation
rRNA Removal - SortMeRNA
Tree Construction - RAxML
Phylogenetic Analysis
Transposable Element Analysis - MITE-Hunter
Blast2GO Annotation
DNA Extraction
Metabolomics with High-Performance Liquid Chromatography (HPLC)
RNA Structure Prediction
Ion Content Detection in vivo and in media
Microscopic Analysis
Metabolite Analysis - Database Search
Statistical Analysis
Tertiary Structure Prediction - SWISS-MODEL
Genomic Analysis with BEDTools
Chromatographic Analysis
Gene Set Enrichment Analysis
Metabolite Analysis - GC-MS and other methods
Genome Assembly and Annotation - Pilon
Gene Set Enrichment Analysis - agriGO
Gene Prediction - OmicsBox
Transcriptome Assembly
Subcellular Prediction - WoLF PSORT
RNA Extraction
Transcriptome Assembly - Quality Assessment
Protein Domain Analysis - SMART
Gene Prediction
Bioinformatics Analysis
Data and Statistical Analysis
Epigenetics
Experimental Techniques
Functional Analysis
Genomics
Metabolomics
Proteomics
Transcriptomics
B
Pointwise Mutual Information
0.40
0.35
0.30
0.25
0.20
0.15
0.10
0.05
DEG Detection - edgeR
Gene Expression level - Normalisation: 0.3432
Gene Set Enrichment Analysis: 0.3538 0.1689
Phenotypic Data Analysis: 0.1329 0.0993 0.0841
RNA Extraction: 0.1054 0.0922 0.0831 0.0910
RNA-Seq Analysis: 0.1670 0.1363 0.1218 0.0732 0.0608
RT-qPCR: 0.1023 0.1187 0.0745 0.0664 0.0637 0.0481
Read Quality Control - FASTX: 0.3869 0.3629 0.2940 0.0799 0.1029 0.1196 0.0797
Sample Preparation: 0.1957 0.1881 0.1246 0.1511 0.1402 0.1038 0.0788 0.1873
Sequence Read Mapping: 0.2867 0.2992 0.1880 0.0564 0.0722 0.0993 0.0613 0.4151 0.1255
DEG Detection - edgeR
Gene Expression level - Normalisation
Gene Set Enrichment Analysis
Phenotypic Data Analysis
RNA Extraction
RNA-Seq Analysis
RT-qPCR
Read Quality Control - FASTX
Sample Preparation
Sequence Read Mapping
C
PMI(A, B) = log prop(A ∩ B) / (prop(A) × prop(B))
prop(A ∩ B) = Proportion of articles where both analyses A and B are present
prop(A) = Proportion of articles where analysis A is present

**Figure 5. LLM-assisted meta-analysis of methods present in the plant stress literature for transcriptomic studies.** (A) All methods/analyses performed across all full-text articles were categorised under 106 supergroups. These supergroups were connected with one another based on the number of articles in which they appeared together. (B) The edge widths between the ten most abundant supergroups are displayed in an annotated heatmap. (C) Edge widths were calculated using the pointwise mutual information (PMI) metric, taking into account the proportion of articles in which each pair of supergroups co-occur in.

Finally, in order to plot a knowledge graph of the analyses used in the plant stress literature, we plotted counts of supergroups as nodes, where the size of the node denotes the number of articles each supergroup was found in, and the edge width represents the pointwise mutual information between the two supergroups (Church & Hanks, 1990), indicating the strength of association based on how often they appear together in articles (Figure 5A). From the annotated heatmap in Figure 5B, we observe patterns like how DEG detection is more strongly associated with other gene expression analyses, indicating that these analyses are more frequently performed together. Similarly, the close relationship between read quality control and sequence read mapping is expected as both are crucial steps in post-processing of sequencing data. Conversely, sequence read mapping and read quality control have significantly less association with more experimental methods such as phenotypic data analysis and RNA extraction. A caveat of the PMI metric is that it enriches for terms which have lower occurrence, resulting in more specialised methods being assigned a higher PMI score.

This information provides an overview of the various methods commonly and uncommonly used by researchers around the world. Indeed, while scientific methods and protocols are relatively well-described in the literature, the connections between them were previously only known intuitively and passed on from teacher to student by word of mouth. These connections have now been visualised in our graph, which will be valuable in improving plant biology specific models. Such endeavours have been rapidly gaining steam in diverse fields such as agriculture (Potamitis, 2023), finance (Araci, 2019), biomedical research (Cai et al., 2023) and materials chemistry (Zheng et al., 2023), and have proven effective at rapidly and accurately distilling the scientific literature to its salient points.

*The application of deep learning and large language models in biology*

The rise of self-supervised deep learning models for DNA and protein sequences, such as ESM-2 (Lin et al., 2023), DNABERT-2 (Zhou et al., 2024) and MSA transformer (Rao et al., 2021), has made it possible to extract previously unknown insights on structure, function and evolution of proteins and DNA sequences (reviewed in (Lam et al., 2024)). In these deep learning models, labelled data is not required, and the model self-learns from raw sequences alone the fundamental properties of the input sequence without any further human guidance. ESM-2 itself is trained on the UniRef dataset, and is primarily trained by masking out certain regions of the protein sequence and training the model to "fill in the gaps" for said masked regions (Lin et al., 2023). This allows the model to fundamentally understand the evolutionary "language" of proteins by inferring amino acid sequences from neighbours, and the model can then be finetuned into specific tasks (Lin et al., 2023). Advances in this field have allowed the accurate gene ontology prediction from protein sequences alone, allowing for elucidation of the three main groupings in gene ontology: biomolecular processes, molecular functions, and cellular components. An example of such a tool is NetGO3, which uses the ESM model to provide state-of-the-art predictions into gene ontology (S. Wang et al., 2023). The accurate prediction of GO terms can, in turn, provide leads that can guide further laboratory validation in a typical gene function prediction approach (Rhee & Mutwil, 2014).

Furthermore, tools such as AlphaFold2 and conformational subsampling of such can provide structural insights into how proteins related to plant stress bind and interact with each other (del Alamo et al., 2022). In essence, by sampling the multiple sequence alignment inputs into AlphaFold2, it is possible to find out what are the possible conformations of a certain protein and better understand the differing functions in different conformations (del Alamo et al., 2022). The latest work has also enabled the *ab initio* prediction of the three-dimensional structure of ligands embedded in proteins, allowing for *de novo* design of binding molecules and cofactors (Krishna et al., 2024). The recent AlphaFold3 can computationally predict the folding and binding conformations of DNA, RNA and common biological ligands as well, which would greatly accelerate the elucidation of molecular interactions and their biological functions (Abramson et al., 2024).

AI in biology is growing at a rapid pace, as exemplified by methods that can design proteins with desired properties by diffusion models (Watson et al., 2023) and predict protein-protein interactions (Bryant et al., 2022). These advances are made possible by the increased ease of generating high-throughput data, more powerful and affordable computer hardware and better AI/ML methods (reviewed in (Lam et al., 2024)). While the above examples provide general predictions pertaining to gene functions, we envision that they can be repurposed to provide insights into stress resilience-related processes.

*Text mining for hidden connections from available data using LLMs*

The burgeoning volume of plant stress research can make it difficult to summarise how the different entities (e.g., genes, metabolites, organs, cell types, organisms) respond to stress and how the different entities are interrelated. While we have demonstrated here only the use of LLMs for the purpose of categorisation in a relatively small subset of the scientific literature (Figure 5), the detection of patterns and trends across these diverse datasets brings out the real strength of using LLMs in such analyses. Previously, we had also demonstrated the text mining capacities of Generative Pre-trained Transformer (GPT) to identify functional relationships between various biological entities such as genes, metabolites and tissues in Arabidopsis (Fo et al., 2023) and yeast (Arulprakasam et al., 2024). Similar approaches were also performed to extract protein-protein or protein-chemical associations from text (Weber et al., 2023). The ability of LLMs to recognise categories of interest when defined in the prompts has also allowed for rapid selection of articles of interest pertaining to the effects of immunotherapy in glioma patients for downstream meta-analyses (Cai et al., 2023), and even to distil out experimental conditions used for metal-organic framework synthesis to build an AI chemistry assistant (Zheng et al., 2023). These approaches harness the inherent strengths of LLMs in their ability to understand natural language to perform parsing and categorisation tasks.

It was shown that ChatGPT has sufficient insight to the scientific corpus to generate a series of pertinent questions in the plant sciences, albeit relatively general in nature (Agathokleous et al., 2023). It was also noted that ChatGPT suffered from ‘plant blindness’ in which its answers to plant-specific questions posed by the user were scientifically inaccurate (Geitmann & Bidhendi, 2023). However, these issues could be addressed by coupling an LLM with a topic specific knowledge-base, that contains distilled information on a topic of interest. For example, Retrieval Augmented-Generation (RAG) comprises a retrieval component that first identifies relevant information from the knowledge base based on the input query (Lewis et al., 2020). The generative component of the LLM then utilises this retrieved information to produce responses that are more accurate and contextually relevant. This approach is particularly beneficial in providing precise and informative answers, especially when the model's inherent knowledge might be outdated or insufficient. A recent development, GeneGPT, is a LLM-based approach which was trained to use the Web APIs of the National Center for Biotechnology Information (NCBI) for answering genomics questions (Jin et al., 2023). While such approaches have not yet been applied to plant stress, an LLM that provides factual information about plant stress would be invaluable.

*Predicting stress resilience and responses from omics data*

Predicting stress resilience that integrates genomic (genotype-to-phenotype (G2P)), transcriptomic, metabolomic and phenomic data could help us understand the molecular basis for stress resilience and engineer more resilient plants (reviewed in (Murmu et al., 2024). As the variation of complex traits is subject to complex regulatory circuits acting on the level of DNA, RNA, proteins and metabolites (Harfouche et al., 2019), genomic signatures that can underpin stress resilience are often difficult to pinpoint, highlighting a need to improve current G2P methods (Cembrowska-Lech et al., 2023; Rico-Chávez et al., 2022). Integration of various types of omics data is known to improve the performance of models. For example, by integrating transcriptomics and metabolomics, yield and phenotypes of maize could be predicted with an increased accuracy (Cruz et al., 2020). Abiotic stress-tolerant crop phenotypes were similarly identified by integrating genomics, transcriptomics and proteomics (Jogaiah et al., 2013). A recent study showed that gene expression in combined stresses (e.g., heat and high light) could be predicted with high accuracy by regression approaches (Tan et al., 2023), showing that stress responses can be predicted even in more complex environments.

While integrating various types of data can improve the performance of predictive models, generating and analysing such data is still difficult and out of reach for most researchers. Fortunately, recent advances in computer science and AI have made it possible to perform new types of analyses that can predict gene expression from DNA sequences. Enformer (a portmanteau of enhancer and transformer) was built to predict gene expression and chromatin states in humans and mice from DNA sequences only (Avsec et al., 2021). While this approach has not been applied to plants yet, such an AI model could be used to predict the expression of stress resilience-conferring genes from DNA sequence, thus enabling new means to identify resilient crops. This goal could also be achieved with AI models that can predict the outcomes of genetic perturbations on gene expression. For example, the Geneformer model trained on DNA and the scGPT model trained on single-cell RNA-sequencing data were able to predict the consequences of genetic perturbations (Theodoris et al., 2023). Such an analysis could be repurposed to plant stress responses, where a model trained on gene expression data from stress experiments could 'learn' to predict stress responses in new experimental settings.

Plant-specific models such as the Genomic Pre-trained Network (GPN) were trained using reference genomes from the Brassicales (including *A. thaliana*) based on a convolutional neural network and were able to identify DNA motifs, various types of genomic regions (intergenic, CDS, introns) and predict the effect of single-nucleotide polymorphisms (Benegas et al., 2023). FloraBERT is a transformer-based model that was trained on the promoter sequences of several hundred species of plants obtained from the Ensembl Plants, RefSeq, and MaizeGD databases and was used to predict and verify against the expression levels of a test set of Zea mays promoter sequences (Levy et al., 2022). AgroNT is a foundational large language model trained on 48 plant reference genomes with a predominant focus on crop plants and was shown to achieve state-of-the-art results for the prediction of genomic elements and gene expression (Mendoza-Revilla et al., 2023). We envision that such approaches will be increasingly used to perform *in silico* mutational experiments to identify stress-resilient plants. While the above approaches have not yet been used to study stress responses, these AI models can potentially be used to identify emerging trends and repeated patterns within the data, providing a richer understanding of how different plant species react to environmental stresses. Such insights are crucial for evolving strategies to boost plant resistance to environmental challenges and for designing new studies in this domain.

*The democratisation of AI and its challenges*

The application of artificial intelligence (AI) in biology holds immense potential, offering revolutionary ways to understand complex biological systems and accelerate research. AI democratises access to sophisticated analytical tools and computational power, enabling researchers from smaller institutions and developing countries to perform advanced analyses without needing extensive resources (reviewed in (Williamson et al., 2023)). Many AI tools and models are available as open-source software, which encourages innovation and allows researchers to build on existing work, driving progress in the field. While AI tools are becoming more accessible, there is still a disparity in computational resources and expertise, which can hinder truly equal access and usage. The widespread availability of AI tools also raises ethical issues related to data privacy, misuse of AI, and the potential for unequal benefits across different socioeconomic groups.

In biology, data can be noisy, incomplete, or inconsistent, which can impair the performance of AI models. The AI approaches often require copious amounts of data that need to be organised and machine-readable (Budach et al., 2022). The inconsistency of proper annotation within databases like the Sequence Read Archive (SRA) and Gene Expression Omnibus (GEO) pose significant challenges for researchers seeking to exploit the abundant wealth of data stored therein (Gonçalves & Musen, 2019). In many instances, essential information such as experimental conditions, plant species and stress types are either inadequately documented or entirely absent, which hampers the ability to categorise and compare studies due to incomplete contextual information. Thus, it is imperative for these databases to prioritise and enforce standardised annotation practices (Brazma et al., 2001; Wilkinson et al., 2016). Although tools have been developed to improve the searchability and visualisation of the inherent metadata (Vaneechoutte & Vandepoele, 2019; Y. Zhu et al., 2013), the underlying issue of missing or incorrect annotation is still not dealt with. Improved annotation would enhance the usability of these valuable resources and expedite scientific progress by enabling researchers to extract meaningful patterns and trends from the available data.

AI systems, particularly generative models, can produce erroneous or "hallucinated" results that do not reflect reality, which can mislead researchers if not carefully validated. These hallucinations may arise from several sources, ranging from inaccuracies in the original training dataset to biases arising during the pre-training of the LLM model (Huang et al., 2023). Retrieval Augmented-Generation (RAG) is a promising strategy used in which the LLM is provided with additional relevant documents retrieved from an external knowledge source, that helps mitigate some of the issues caused by 'hallucinations' (Guu et al., 2020; Lewis et al., 2020). AI models may also suffer from "catastrophic forgetting", where they lose previously learned information when new data is introduced (French, 1999). Keeping models updated with the latest data without losing previous knowledge is a challenge. Recent advances in addressing this issue have drawn inspiration from biology where physical modifications to excitatory synapses in the brain were shown to correlate with the acquisition of new knowledge or skills, and protect against forgetting (Yang et al., 2009). Computer scientists have mimicked this concept by developing an algorithm termed elastic weight consolidation (EWC), which slows down learning on certain weights based on how important they are to previously seen tasks (Kirkpatrick et al., 2017).

Many AI models, especially deep learning networks, function as "black boxes," providing little insight into how decisions are made. This lack of interpretability can be problematic in understanding biological mechanisms and ensuring trust in AI-driven conclusions (Ali et al., 2023). Reproducing results generated by AI models can be challenging due to variability in data, model parameters, and computational environments (reviewed in (Williamson et al., 2023), which complicates the validation and verification of scientific findings. This challenge is being tackled head on by the field of eXplainable Artificial Intelligence (XAI), which has made significant strides in developing methods or alternative models to provide greater interpretability and hence confidence in the model output for a large number of real-world applications (reviewed in (Saranya & Subhashini, 2023)). 'White-box' models, which as the term suggests, allows for the interpretation of the logic and processes occurring between input and output. The models are typically designed using linear regression, decision tree, and rule-based models, which excel at performing single tasks, but fall behind current deep-learning models in the context of more general artificial intelligence. Most interpretability methods for explaining deep learning models apply to image classification and feature the concept of saliency maps, which highlights the relative 'importance' of the different image regions in producing the output (reviewed in (Linardatos et al., 2021)). Other approaches to understand the logic behind the relative weights and connections generated within the model involve the use of adversarial examples to invoke errors such as misclassifying an image by applying a single minute perturbation in the sample input (Szegedy et al., 2014).

**Conclusions and perspectives**

Plant stress research is essential for us to develop crops that can thrive in more challenging environments. Biological systems are robust and reproducible, despite comprising millions of components interacting in ways that have evolved over billions of years of selection. The resulting systems are marvellously complex and are beyond current human comprehension (Batzoglou, 2023). To be able to model biological systems, we have to resort to simplistic rules that might result in digestible but incomplete narratives. However, the rapidly accumulating biological data and improving AI methods could be used to connect the different types of data and identify yet unknown and possibly unimaginable patterns present in biological systems.

The democratisation of AI-assisted tools heralds a revolution in biology and medicine. Previously the domain of specialised professions, LLM-based AI agents such as ChatGPT, Mistral, Claude, Gemini and others, have virtually taken over the world in diverse industries from agriculture (Harfouche et al., 2019) to zoology (Love et al., 2024). From decoding the secrets of the plant genome to understanding animal language, it seems that the development of AI is opening up completely new vistas in previously intractable fields, and the only limits are of our imagination and creativity in finding new problems to solve. The harnessing of this new technological wave promises significant acceleration of our efforts to understand more about plants and how they respond to diverse stresses across the plant kingdom. These developments have arrived at a critical juncture, where humanity is faced with the looming threat of climate change, and a rapidly growing population.

Addressing the impending challenges of climate change and global food security demands a concerted and collaborative effort from the global scientific community. Scientists play a crucial role in developing innovative, sustainable and resilient solutions to mitigate the impacts of climate change on agriculture and enhance global food security. The widening gap between richer and poorer countries in their ability to address climate change and food security requires a collective commitment to global equity. International collaborations, technology transfer and fair distribution of resources can contribute to narrowing this gap, ensuring that vulnerable populations have the means to adapt to changing climates and secure their food supply. As scientists contribute their expertise and advocate for inclusive policies, they play a pivotal role in fostering a more sustainable and equitable future for global food security.

**Table 1. Summary of databases used in plant stress research.** Each database is annotated with the type of data it contains, its area of focus, the type of species from which the data was obtained, and its associated reference(s).

| Database | Data type | Focus area | Species type | References |
|---|---|---|---|---|
| Stress Knowledge Map | Knowledge graphs. Protein - protein interactions Protein - DNA interactions, smallRNA - transcript interactions Enzymatic transformations of metabolites | Heat, Drought, Flooding, Extra- and intracellular pathogens, Herbivory | *Arabidopsis thaliana*, *Solanum tuberosum* | Bleker et al., 2024 |
| Plant stress RNA-seq Nexus (PSRN) | Transcriptomic, long noncoding RNAs (lncRNAs) | ABA, Darkness, Cold, Pathogen, Ozone, Nutrition, Dehydration, Heat, Light, NaOH, PEG-8000 | *Arabidopsis thaliana*, *Chlamydomonas reinhardtii*, *Glycine max*, *Manihot esculenta*, *Oryza sativa* indica, *Oryza sativa* Japonica, *Panicum virgatum*, *Populus tremuloides*, *Solanum lycopersicum*, *Sorghum bicolor*, *Triticum aestivum*, *Vitis vinifera* | Li et al., 2018 |
| Plant Environmental stress transcript database | Transcriptomic | Heat, Cold, Dehydration, Salt | Wheat, Maize, Rice, Barley, Sorghum, Pearl millet, Rye, Arabidopsis, Common bean, Tomato, Soybean, Cowpea, Groundnut, Potato, Chickpea, Medicago | Balaji et al., 2006 |
| Plant Resistance Genes database (PRGdb 4.0) | Transcriptomic | Plant disease resistance | *Solanum lycopersicum*, *Oryza sativa*, *Triticum aestivum*, *Vitis vinifera* and *Arabidopsis thaliana* | Calle García et al., 2022 |
| DroughtDB | Transcriptomic | Drought stress response | Arabidopsis , rice, sorghum, maize, Brachypodium, tomato, barley, rye and *Aegilops tauschii* | Alter et al., 2015 |
| Stress Responsive Transcription Factor Database (STIFDB V2.0 ) | Transcriptomic, Transcription Factor Binding Site | Bacteria, ABA, drought, cold, salinity, osmotic, wounding, dehydration, UV-B, high light, heat, heavy metals | *Arabidopsis thaliana* and *Oryza sativa* L. | Naika et al., 2013 |
| CerealESTDb | Transcriptomic | ABA, cold, drought, heat and salt | Maize, Rice, Sorghum, Wheat | Kumar et al., 2022 |
| HEATSTER | Genomic | Heat stress transcription factors | 65 species including 1 Phaeophyta, 3 Rhodophyta, and 61 species of Viridiplantae (5 Chlorophyta, 1 Bryophyta, 1 Lycopodiidae, 1 basal Magnoliophyta, 3 Gymnosperms, 15 Monocotyledons, and 35 Eudicotyledons) | Berz et al., 2019 |
| Rice Stress-Resistant SNP database | Genomic SNPs specific to biotic and abiotic stress-resistant | Cold, Salt, Rice blast fungus, Heat, Bacterial leaf blight, Flood, Alkali, Bacterial stripe virus, Brown planthopper, White-backed planthopper, Bacterial sheath blight, Gall midge pest, Bacterial planthopper | Rice | Tareke Woldegiorgis et al., 2019 |
| PncStress | ncRNAs microRNAs, long non-coding RNAs, circular RNAs | 48 biotic and 91 abiotic stresses | 114 species | Wu et al., 2020 |
| CropCircDB | circular RNAs | Drought, Cold, Salt | Maize and Rice | Wang et al., 2019 |
| PlaASDB | Alternative splicing events | Salt, Heat, Hormone, Wounding, Drought, Osmotic, Dark, Chemical, Nutrient, Pathogen | Arabidopsis and Rice | Guo et al., 2023 |